# Nanocavity Induced Light Concentration for Energy Efficient Heat Assisted Magnetic Recording Media

Chenhua Deng[a,b,c,1], Haomin Song[d,1], James Parry[b], Yihao Liu[e], Shuli He[e], Xiaohong Xu[a], Qiaoqiang Gan[d,*], Hao Zeng[b,*]

[1]These two authors contributed equally to this work.

[a]College of Chemistry and Materials Science, Shanxi Normal University, Linfen 041004, China

[b]Department of Physics, University at Buffalo, SUNY, Buffalo, New York 14260, USA

[c]College of Chemistry, Taiyuan Normal University, Jinzhong 030619, China

[d]Department of Electrical Engineering, University at Buffalo, SUNY, Buffalo, New York 14260, USA

[e]Department of Physics, Capital Normal University, Beijing 100048, China

E-mail: qqgan@buffalo.edu (Q. Gan); haozeng@buffalo.edu (H. Zeng)

**Abstract:**

Enhancing light absorption in the recording media layer can improve the energy efficiency and prolong the device lifetime in heat assisted magnetic recording (HAMR). In this work, we report the design and implementation of a resonant nanocavity structure to enhance the light-matter interaction within an ultrathin FePt layer. In a Ag/SiO$_2$/FePt trilayer structure, the thickness of the dielectric SiO$_2$ layer is systematically tuned to reach maximum light absorption at the wavelength of 830 nm. In the optimized structure, the light reflection is reduced by more than 50%. This results in effective laser heating of the FePt layer, as imaged by an infrared camera. The scheme is highly scalable for thinner FePt layers and shorter wavelengths to be used in future HAMR technologies.

# 1. Introduction

$L1_0$ structured Iron Platinum (FePt) thin films are candidates for recording media of the next generation hard disk drives with high areal density due to their large magnetocrystalline anisotropy, making them thermally stable even for very small grains [1-3]. However, their large coercivity tends to exceed the writability of the recording head. Heat-assisted magnetic recording (HAMR) can lower the coercivity of the media momentarily using laser light heating and allow writing on the heated area of the high anisotropy films [4-6]. Therefore, HAMR has been considered as a promising route to combat this issue [7-9] and increase the areal density of hard disk drives by as much as 100 folds [10-11]. Current HAMR system delivers the incident light power of a laser diode to the recording media using a near field transducer (NFT) [12-13]. An efficient HAMR process requires the major part of the energy to be absorbed by the magnetic layer [14-16]. However the ultrathin metallic FePt films show insufficient light absorption, resulting from the thicknesses orders of magnitude smaller than the wavelength of the incident light. On one hand, a significant portion of the light is reflected back by the FePt layer and may overheat the NFT, leading to head fatigue and eventually head failure. On the other hand, the light transmitted through the FePt film may complicate the design of heat dissipation layer and affect the temperature control in the HAMR process. For the HAMR technology to be commercially viable, it is highly desirable to enhance the light absorption of the ultrathin FePt layer to improve the energy efficiency and device lifetime.

A potential approach is to utilize interference resonance to enhance the light-matter interaction within the FePt layer. Traditional Fabry-Perot-type of interference has been widely employed in thin-film structures based on dielectric and semiconductor cavities to develop a variety of applications ranging from thin-film optical filters [17] to anti-reflection coating layers [18-19]. Recently, interference in absorbing systems received increased interests [20]. For instance, M. A. Kats *et. al.* [21] obtained high optical absorption in 7~25 nm Ge films by coating it on a Au film. Due to the additional large phase shift occurred at the Ge/Au interface, the thickness of the absorbing layer can be reduced to be much smaller than a quarter wavelength to meet the destructive interference condition. However, the spectral position of the resonance is determined by the thickness of the absorbing layer, i.e., one has to increase the thickness to realize resonances at longer wavelengths. To overcome this thickness limitation, more recently, Song *et. al.* [22] achieved a tunable enhanced absorption by introducing a lossless dielectric layer between the metal reflector and the absorbing layer. The desired destructive interference can then be adjusted by tuning the thickness of the lossless layer, rather than increasing the thickness of the absorbing layer. Importantly, one can select low loss metal as the bottom reflector material to minimize the optical attenuation in this layer and, therefore, enhance the exclusive optical absorption within the ultrathin absorbing layer, enabling the development of more energy efficient optoelectronic devices (e.g. ultra-thin Ge photodetectors [23-24]).

Inspired by this nanocavity-enhanced light-matter interaction mechanism, here we propose the design and fabrication of a metal-dielectric-metal resonant nanocavity structure to concentrate light and heat within ultrathin FePt films for HAMR applications. We show that the spectral response of the structure can be tuned across the visible to infrared range by adjusting the thickness of the dielectric spacer layer. We further show that a FePt layer as thin as 10 nm can absorb 87% of the incident light at 830 nm (i.e., the wavelength of the diode lasers used by present HAMR technology) and generate heat effectively. In previous work reporting enhanced light absorption for energy harvesting and conversion applications [25-27], the conversion from light to heat is usually undesirable. In this work, however, the nanocavity structure promises to confine light absorption and enhance energy dissipation into heat in the absorbing thin film to change the magnetic properties of the recording media. This design can be implemented in future HAMR system design to reduce energy consumption and premature head failure.

## 2. Experimental section
### 2.1 Fabrication

To validate the principle, a nanocavity structure consisting of $Ag/SiO_2/FePt$ trilayers was deposited on Si substrates at room temperature using sputtering. The sputtering was processed with an Ar gas pressure of $7.5 \times 10^{-3}$ Torr with the base pressure in the low $10^{-7}$ Torr. The FePt layer was deposited using an $Fe_{55}Pt_{45}$ composite target. The thicknesses of Ag and FePt layers were fixed at 80 and 10 nm, respectively, with the thickness of $SiO_2$ tuned from 80 to 160 nm. All samples were annealed at 550 °C in vacuum for 30 min to transform the FePt layer into the hard magnetic $L1_0$ phase. The light absorption of the FePt layer can be manipulated from ultraviolet to visible and eventually to near-infrared region simply by tuning the thickness of the $SiO_2$ layer in this trilayer system.

**2.2 Measurements and characterizations**

UV-vis optical spectra was collected using an optical measurement system consisted of a spectrometer (QE65000, Ocean Optics Inc.), an optical fiber probe, a light source (DH-2000-BAL, Ocean Optics Inc.), and a detector. The detection range of the spectrometer is 200-1000 nm, the reflected light was analyzed using computer spectro-analysis program (Spectrasuit, Ocean Optics Inc.) connected to the spectrometer [28]. We analyzed the spectral position of the reflectivity valley, which corresponds to the absorption maximum. The magnetic properties were characterized by a superconducting quantum interference device (SQUID) magnetometer with the maximum magnetic field of 5 T and can detect moment around $10^{-8}$ emu. The top and cross-sectional morphologies were obtained using scanning electron microscope (SEM) (JSM，7500F) with the largest acceleration voltage of 30 KV and distinguishability of 1 nm. The laser induced heating was measured by an infrared camera (JENOPTIK VarioCAM), while the sample was illuminated with a 830 nm laser with an output power of 400 mW and spot size of 4 mm. The dispersive optical constants of a 10-nm-thick FePt film were characterized using a spectroscopic ellipsometer (J. A. Woollam, VASE), and were adopted in the design of super absorber using rigorous coupled-wave analysis (RCWA) method in the following section.

**3. Results and discussion**

**3.1 Design of nanocavity super absorber**

Due to the ultra-thin thickness of the FePt film, the optical absorption within this layer is relatively weak. According to our numerical modeling using rigorous coupled-wave analysis (i.e., RCWA model of Rsoft), a 10 nm-thick FePt film on a glass substrate (Figure 1a) can only absorb 39.8 % of the incident light at 830 nm. A significant portion of the light is either reflected (37.1%) or transmitted (23.1%) (see black and blue curve in Figure 1b, respectively). This numerical modeling is confirmed by the measured reflection and transmission spectra for a 10-nm-thick FePt film on a glass substrate, with the measured reflectance and transmission being 36.6% and 21.3%, respectively (see black and blue dots in Figure 1c, respectively), consistent with the simulation results in Fig. 1b. The reflected energy will heat the writing head with a NFT, and resulting in head fatigue and failure. Therefore, reflection should be minimized to prolong the lifetime of the transducer head.

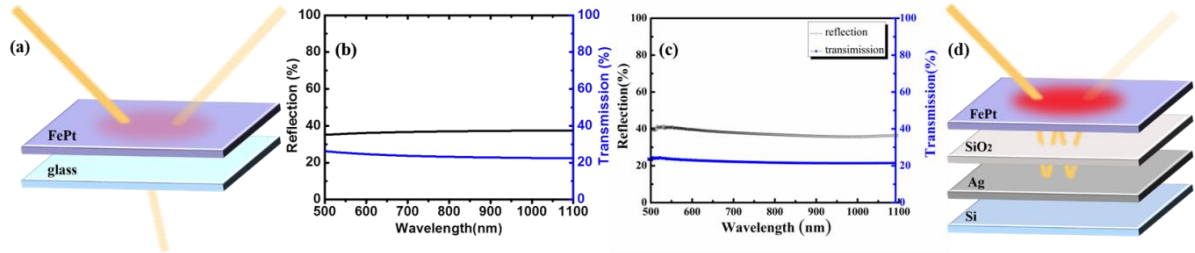

**Figure 1.** (a) Schematic illustrations of light-matter interaction in FePt on a glass substrate; (b) The simulated reflection (black) and transmission (blue) spectra of 10 nm-thick FePt film on $SiO_2$ substrate; (c) The measured reflection (black) and transmission (blue) spectra of a 10 nm FePt film on glass substrate; (d) Schematic illustrations of light-matter interaction in $Ag/SiO_2/FePt$ trilayer structure on a glass substrate, demonstrating enhanced light absorption.

To enhance the light absorption of the FePt layer at 830 nm, we designed a metal-dielectric-metal nanocavity structure (Figure 1d), employing a silver (Ag) back reflector to reduce the energy loss in the FePt film in the 800-900 nm wavelength region. A $SiO_2$ film was selected as the phase compensation layer to manipulate the resonant absorption of the nanocavity. $SiO_2$ is also a typical substrate for FePt growth to produce perpendicularly orientated FePt grains without epitaxy [29-30], which is desired for data-storage applications. The absorption resonance in the FePt layer can be manipulated by controlling the thickness of the $SiO_2$ layer. Therefore, we fixed the thickness of the Ag reflector and FePt absorptive layer at 80 and 10 nm, respectively, and tuned the thickness of the $SiO_2$ dielectric layer. As shown in Figure 2a, we first modeled the total absorption (*A*) of the system as a function of the $SiO_2$ thickness. One can see a redshift of the resonant peaks and valleys as the thickness of the $SiO_2$ layer increases, indicating the spectral tunability of the trilayer structure. For example, the absorption valley shifts from 450 to 560 nm by adjusting the $SiO_2$ thickness from 120 to 160 nm. Remarkably, during the redshift, the total absorption of the entire system could be increased to 81.6% at 830 nm when the thickness of $SiO_2$ layer was tuned to 160 nm. Based on Figure 1a, the reflection (*R*) of the system can also be determined by the expression $R = 1 - A - T$. Here, *T* is the transmission, which is zero due to the bottom 80-nm-thick Ag layer. Therefore, compared with the reflection of the bilayer 10-nm-FePt/glass at 830 nm (i.e., 36.6%), the reflection of the trilayer system can be suppressed remarkably by changing the $SiO_2$ thickness. For example, when the thickness is tuned to 160 nm, the reflection can be reduced to be as low as 18.4%, a ~ 50% reduction. The significantly reduced reflection is beneficial for reducing the heating of the NFT. However, Figure 2a only demonstrates the enhanced total absorption of the entire system. To achieve the enhanced heating effect of the FePt layer, most of the light should be absorbed in this recording layer. Therefore, we further modeled the exclusive absorption of the FePt layer in Figure 2b, confirming that the major light dissipation occurs in the top recording medium. The absorption of the Ag layer can be estimated by calculating the difference between Figure 2a and b, showing that it is close to a perfect reflector at 830 nm. To further visualize the enhanced absorption effect, the absorption and the optical field distribution of the trilayer structure with a $SiO_2$ thickness of 160 nm are modeled using finite element method and rigorous coupled-wave analysis, respectively, as shown in Figure 2c (at 830 nm). Due to the nanocavity interferometer mechanism, the optical field was confined within the tri-layer system, as shown by the white dashed line. One can see that most of the energy was dissipated in the top FePt layer and the loss in the bottom Ag layer is negligible, well satisfying the requirement for HAMR applications. Note that when the thickness of the $SiO_2$ layer is in the range of 120 to 160 nm, the exclusive absorption over 70% at 830 nm can be maintained within the 10 nm-thick FePt layer (see Figure 2b), demonstrating good tolerance of the $SiO_2$ thickness in real fabrication.

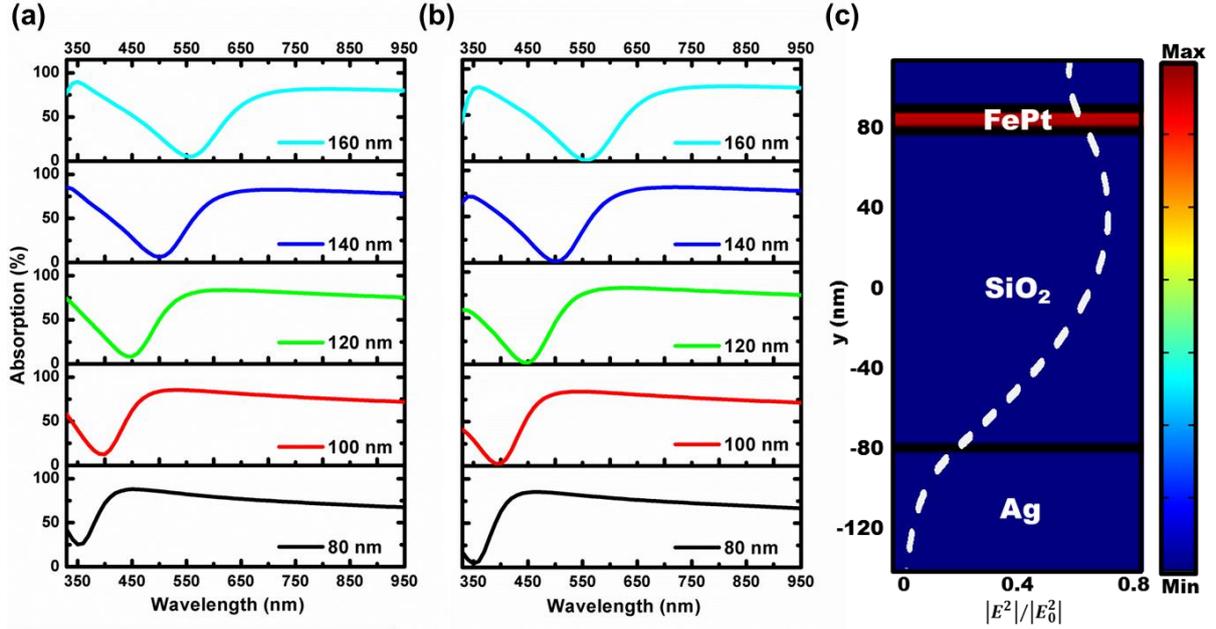

**Figure 2.** Simulated (a) total absorption spectra of the trilayer system and (b) exclusive absorption spectra in the 10-nm-thick-FePt layer on Ag/SiO$_2$ substrates, as a function of the SiO$_2$ thickness. (c) The absorption distribution in an Ag/160-nm-thick-SiO$_2$/10-nm-thick-FePt trilayer structure at 830 nm.

## 3.2 Experimental implementation

To confirm the simulation results, we fabricated a series of Ag/SiO$_2$/FePt trilayer structures with the SiO$_2$ thickness ranging from 80-160 nm. The spectral position of the absorption band (corresponding to the reflectance minimum) is shown in Figure 3a. One can see that the reflection valley red-shifts as the SiO$_2$ thickness increases and approaches 830 nm when the thickness of SiO$_2$ is around 120-140 nm, agreeing reasonably well with the theoretical prediction. For example, the reflection is 13% at 830 nm at the spacer layer thickness of 140 nm, showing a strong suppression of the reflection at the FePt/air interface, which is highly desired for HAMR application. Figure 3b shows the photographs of Ag/SiO$_2$/FePt samples under white light illumination, with SiO$_2$ thicknesses ranging from 80 to 160 nm. One can see a range of vivid colors due to the tunable reflection spectra depending on the thickness of the dielectric layer.

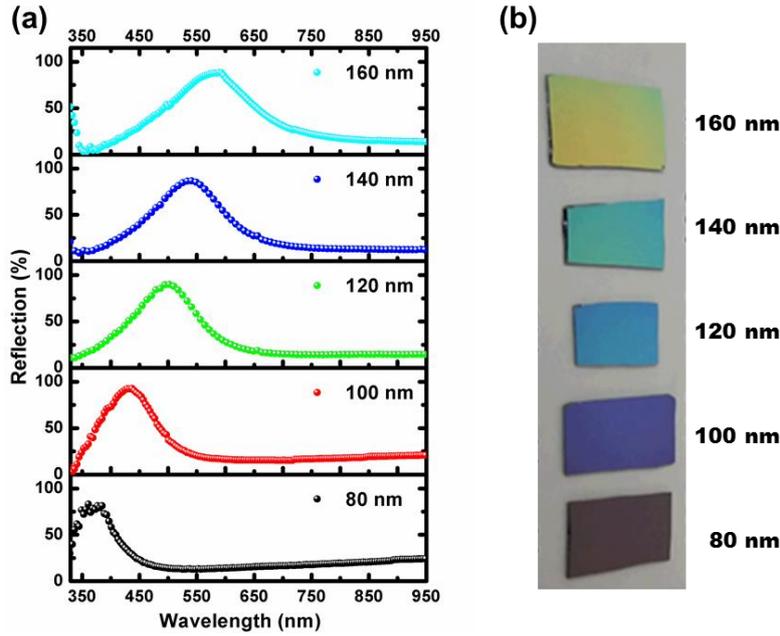

**Figure 3.** (a) Experimental reflection spectra of Ag/SiO$_2$/FePt structures with thickness of SiO$_2$ ranging from 80 to 160 nm, and (b) the corresponding photographs of the samples.

The above experimental results show that by optimizing the thickness of the dielectric space layer, one can confine most of the light absorption at 830 nm within the thin FePt layer. Such light absorption should lead to enhanced energy dissipation and heating of the metallic FePt layer. To verify this, an infrared camera was used to measure the surface temperature of the nanocavity structure under laser illumination at 830 nm. To demonstrate the enhanced light-matter interaction, two samples with different spacer layer thicknesses of 140 nm and 220 nm were selected. According to the measured absorption spectra, the 140 nm sample shows the absorption of 87% at 830 nm while the 220 nm one shows only 2.7% (see Figure S1 in Supplementary Information). The infrared images of the 140 nm and 220 nm samples mounted on the rack are shown in the left panels of Figure 4a and b, respectively. The infrared images of the center spots illuminated by laser at different times are shown in the right panels. The spot size is 4 mm in diameter with the intensity of ~ 2500 mW/cm$^2$. One can see from Figure 4a that the temperature of the first sample with a 140-nm-thick SiO$_2$ layer increased rapidly upon light illumination, and reached a steady state at the temperature of 50 °C after 55-second. While for the sample with the 220-nm-thick SiO$_2$ film (Figure 4b), there is barely detectable temperature change after the 55-second continuous illumination since most of the light is reflected. This experiment clearly demonstrated the enhanced heating effect of the cavity structure.

The enhanced heating effect of the nanocavity structure can be understood by theoretically estimating the temperature rise based on the obtained experimental results. After the samples reach thermal equilibrium under illumination of CW laser, the absorbed laser power, $I_{abs}$, equals to the heat dissipation rate, $q_{dis}$. Here, $I_{abs} = \alpha I_{laser}$, and $\alpha$ is the optical absorption coefficient (i.e., *α = 87%* and *α = 2.7%* for samples for spacer layer thicknesses of 140 nm and 220 nm, respectively) and $I_{laser}$ is the intensity of the CW laser (i.e., 2500 mW/cm$^2$). The heat dissipation results from radiation, convection and conduction, i.e., $q_{dis} = \varepsilon\sigma(T_2^4 - T_1^4) + h(T_2 - T_1)$. Here, $\varepsilon$ is the emissivity taken to be 1 for a first order approximation, $\sigma$ the Stefan-Boltzmann constant, $T_2$ the temperature of the sample at equilibrium, $T_1 = 20$ °C the ambient temperature, and $h$ the heat transfer coefficient due to conduction and convection. Considering the experimental result of the sample with a 140-nm-thick SiO$_2$ layer. i.e., *$T_2$ = 50* °C, we can obtain *h = 718.4 W/(m$^2$·K)*. For simplicity, we assume that the heat transfer coefficient is sample temperature and structure independent. With this heat transfer coefficient, the calculated temperature of the sample with a 220-nm-thick SiO$_2$ layer will reach 20.9 °C under illumination, which agrees well with the experiment result shown in Figure 4b. That is, the 220 nm sample can barely be heated up. We can further estimate the laser power needed to heat the samples up to a steady state temperature of 477 °C, which is the Curie temperature of FePt. The temperature of the two samples as a function of the laser power is shown in Figure 4c. One can see that to achieve the Curie temperature, the required laser intensity is 3.97×10$^4$ mW/cm$^2$ and 1.28×10$^6$ mW/cm$^2$ for the two samples with different spacer layer thicknesses of 140 nm and 220 nm, respectively. This clearly demonstrates that effective heating can only be achieved in optimized nanocavity structure with maximized light absorption and minimized reflection. It should be noted that the laser power calculated is likely a vast overestimate, as the infrared camera used does not have the temporal resolution to measure the instantaneous local temperature of the heated spot, which is expected to be significantly higher than the steady state temperature.

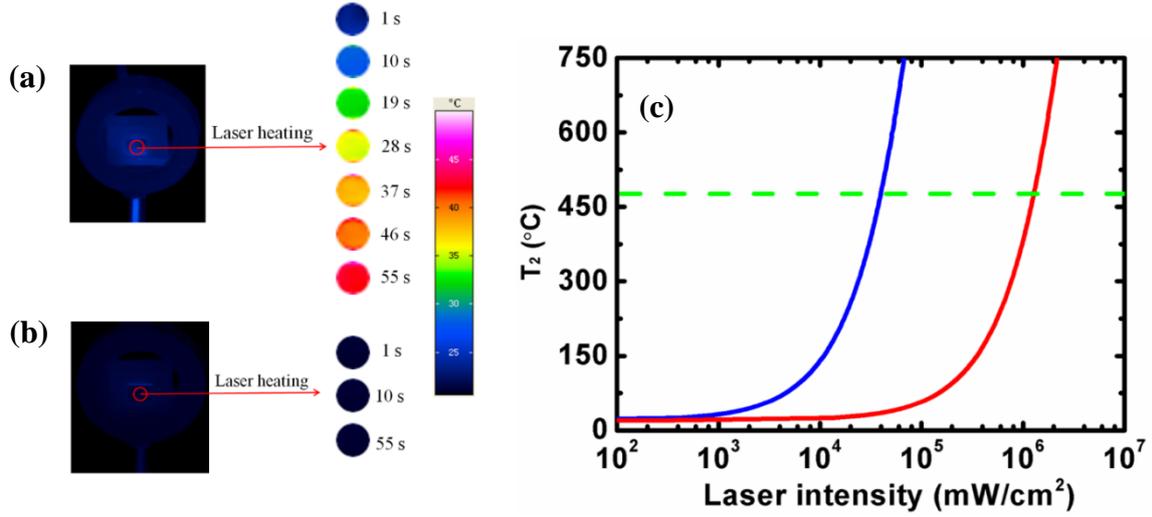

**Figure 4.** The laser heating results of (a) Ag/140-nm-thick-SiO$_2$/FePt before and after 1, 10, 19, 28, 37, 46, 55s laser irritation, (b) Ag/220-nm-thick-SiO$_2$/FePt before and after 1, 10, 55s laser irritation. (c) Calculated temperature of two samples with spacer layer thicknesses of 140 nm (blue solid curve) and 220 nm (red solid curve), respectively. The Curie temperature of FePt (477 °C) is indicated by the green dashed curve.

More importantly, as the areal density in a HARM system increases, thinner FePt films and laser light at shorter wavelengths need to be used. To investigate the suitability of the resonant nanocavity scheme for future more challenging HAMR applications, the absorption characteristics of the trilayer with a 5 nm thick FePt absorber layer is simulated. As shown in Figure 5, the exclusive absorption in the 5-nm-thick-FePt layer can reach 97.7% at 830 nm and 96.6% at 500 nm when it is placed on Ag/160-nm-thick-SiO$_2$ and Ag/80-nm-thick-SiO$_2$, respectively. Therefore, this trilayer nanocavity structure is highly scalable for thinner FePt layers and shorter wavelengths, showing great potential for future HARM system to satisfy the ever increasing demand in areal density.

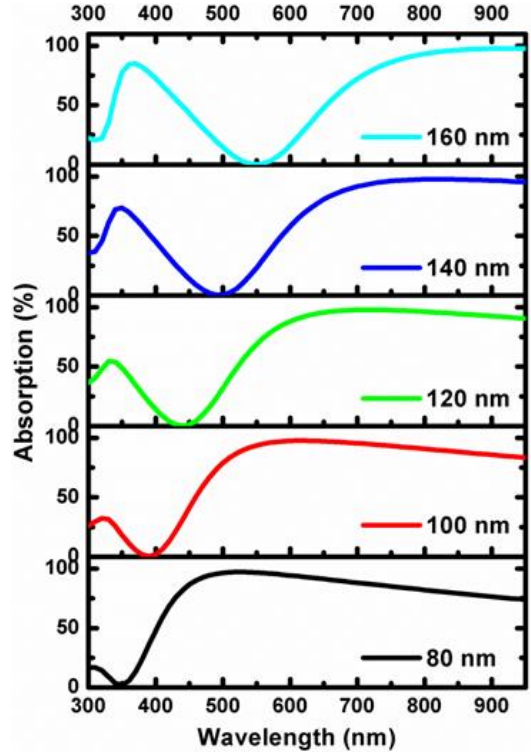

**Figure 5.** Simulated exclusive absorption spectra in the 5-nm-thick-FePt layer on Ag/SiO$_2$ substrates, as a function of the SiO$_2$ thickness.

## 3.3 Morphology and magnetic properties

Finally, it is also important to show that the magnetic and structural properties of the FePt film employed in the nanocavity structure are suitable for magnetic recording applications. To demonstrate this feature, the magnetic property of the 10 nm FePt film in the nanocavity was characterized by a SQUID magnetometer. The films were annealed in vacuum at 550 °C for 30 min to realize the phase transition from disordered fcc to $L1_0$ structure [29-30]. The magnetic hysteresis loop measured at room temperature is shown in Figure 6a. It can be seen that the hard $L1_0$ phase is obtained, with a coercivity of 6.9 kOe. This suggests that the nanocavity structure does not affect the phase transition and magnetic performance of FePt upon annealing. To realize a high storage density and reading stability, the distance fluctuation between the head and the recording medium should be controlled within tens of nanometers (*e.g.* the head-to-medium distance is approximately 15 nm in a recently reported experiment [5]). Hence the surface roughness of the FePt layer should be sufficiently small to avoid the damage of the head due to potential direct contact or friction with the medium. As shown in Figure 6b, the surface of the FePt layer on an Ag/140-nm-thick-SiO$_2$ substrate was characterized using AFM. The surface roughness is ~0.7 nm, which is sufficiently small for the potential application for HAMR. The profile of a typical trilayer system was also investigated using scanning electron microscope (SEM) in Figures 6c and 6d, showing the smooth interfaces on top of a 140-nm-thick SiO$_2$ layer before and after annealing.

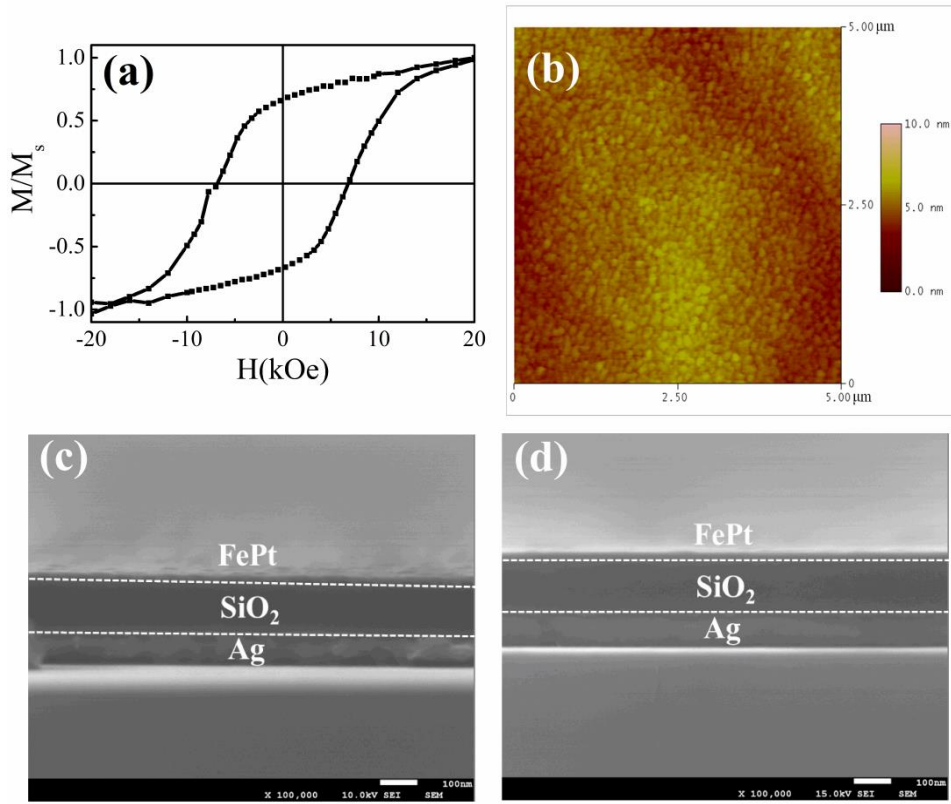

**Figure 6.** (a) Hysteresis loop measurement by SQUID at room temperature of 10 nm FePt film deposited on Si substrate. (b) The AFM image of the sample surface. (c-d) SEM images of the profile of the Ag/140-nm-thick-SiO$_2$/FePt trilayer structures (c) before and (d) after annealing.

Our work is a proof-of-principle demonstration of interference enhanced light absorption and induced heating in the magnetic layer based on nanocavities. Present HARM media design already incorporates a dielectric layer (MgO) and metallic heat sink layer underneath the FePt film. Thus our scheme does not lead to a significant alteration of the present media structure. By tuning the thickness and composition of these layers, it is possible to enhance the light absorption and heating efficiency without compromising the magnetic performance of the media.

## 4. Conclusion

In conclusion, we proposed and demonstrated a resonant nanocavity structure consisting of Ag/SiO$_2$/FePt trilayers to enhance the light-matter interactions in the FePt layer. By tuning the dielectric layer thickness, a 10 nm FePt layer can absorb up to 87% of the incident light at the wavelength of 830 nm. The absorbed light energy can be converted into heat to effectively raise the temperature of the FePt film. This scheme can be incorporated into the design of heat assisted magnetic recording media to reduce power consumption and avoid premature head failure. The design scheme is scalable for thinner FePt layers and shorter wavelengths, which may be useful for extending HAMR technology to higher recording density.

**Acknowledgements**

We thank the financial support from the US National Science Foundation (Grant No. CMMI1562057, ECCS1507312, CBET1510121 and MRI1229208), and the National Natural Science Foundation of China (Grant No. 61434002, 11611540333, 51571146, 51771124).

# Supporting Information

# Nanocavity Induced Light Concentration for Energy Efficient Heat Assisted Magnetic Recording Media

Chenhua Deng[a,b,c,1], Haomin Song[d,1], James Parry[c], Yihao Liu[e], Shuli He[e], Xiaohong Xu[a], Qiaoqiang Gan[d,*], Hao Zeng[c,*]

[1]These two authors contributed equally to this work.

[a]College of Chemistry and Materials Science, Shanxi Normal University, Linfen 041004, China

[b]College of Chemistry, Taiyuan Normal University, Jinzhong 030619, China

[c]Department of Physics, University at Buffalo, SUNY, Buffalo, New York 14260, USA

[d]Department of Electrical Engineering, University at Buffalo, SUNY, Buffalo, New York 14260, USA

[e]Department of Physics, Capital Normal University, Beijing 100048, China

E-mail: qqgan@buffalo.edu (Q. Gan); haozeng@buffalo.edu (H. Zeng)

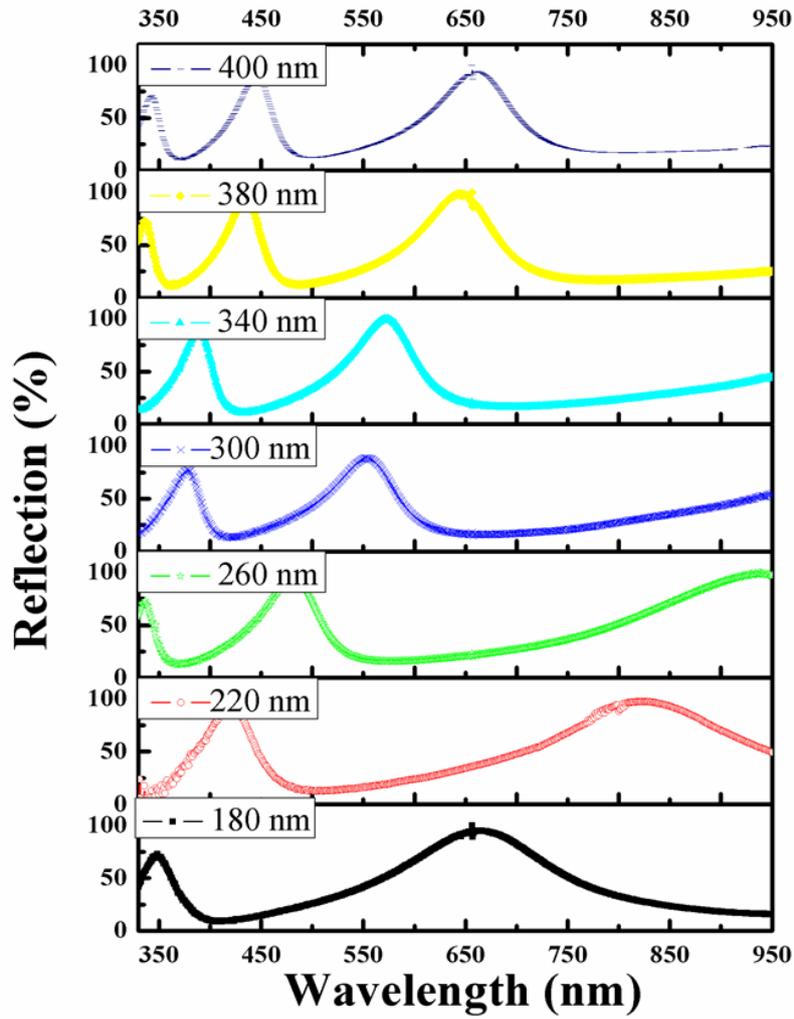

Figure S1. Optical reflection spectra of Ag/SiO$_2$/FePt structures with thickness of SiO$_2$ range from 180 to 400 nm.

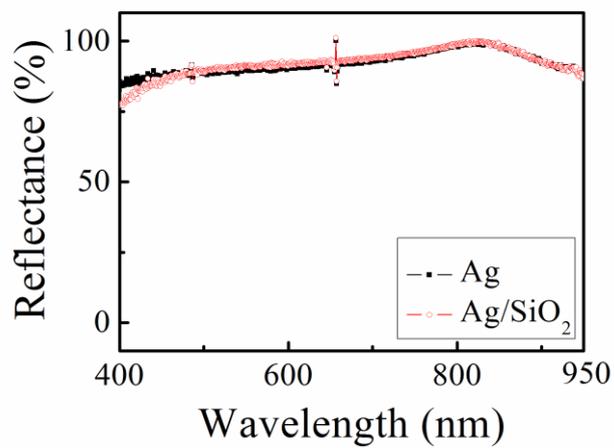

Figure S2. Optical reflection spectra of 80-nm-thick-Ag (black) and 80-nm-thick-Ag/160-nm-thick-SiO$_2$ films (red).

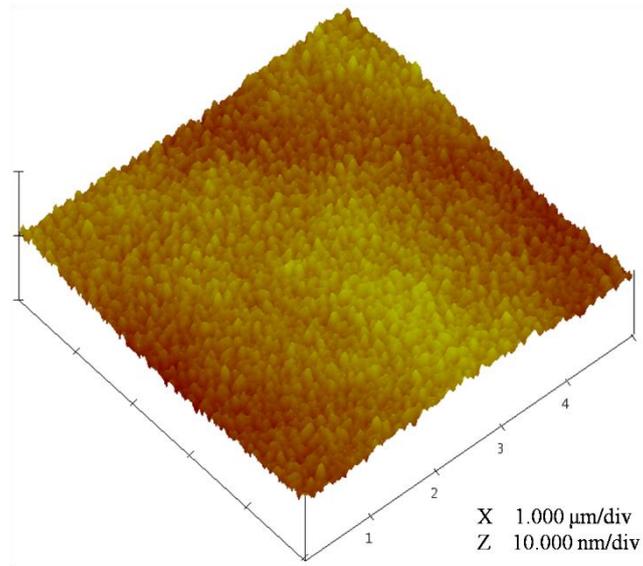

Figure S3. The AFM 3D image of the Ag/140-nm-thick-SiO$_2$/FePt.

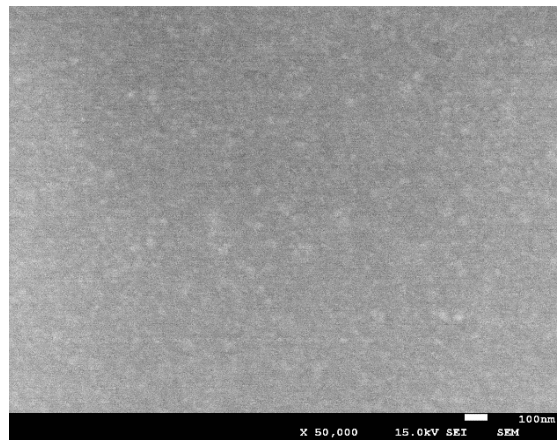

Figure S4. The SEM image of the surface of Ag/140-nm-thick-SiO$_2$/FePt trilayer structure.

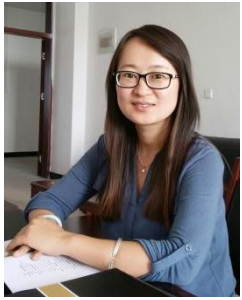

**Chenhua Deng** received her Ph.D. degree from Shanxi Normal University in 2016. She is currently a lecturer at Taiyuan Normal University. Her current research interest includes magnetic recording media, semiconductor spintronic, Chemistry and physics of nanoscale materials.

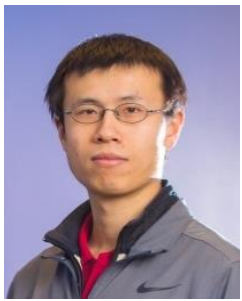

**Haomin Song** received his Ph.D. degree from University at Buffalo in 2018. His current research interests include nanomanufacturing, energy harvesting, conversion and nanophotonics. His research publications include ~30 technical papers. He is currently the CTO of Sunny Clean Water LLC.

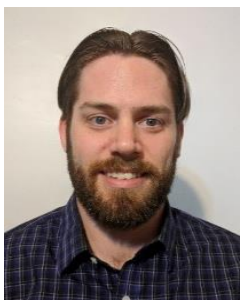

**James P. Parry** is an assistant professor in the Department of Biology and Mathematics at D'Youville College. He received his Ph.D. in Physics from the State University of New York at Buffalo in 2017. His current research interests are energy harvesting and material property modification using optical effects in thin films.

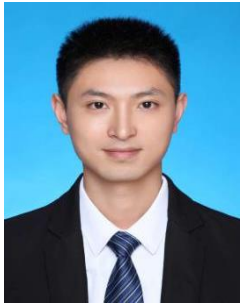

**Yihao Liu** received his M.M. in orthopaedics from the Jiangsu University in 2017. He joined the Chinese PLA General Hospital in 2017. His research interests are in biomaterials, bone tumor and Magnetic hyperthermia.

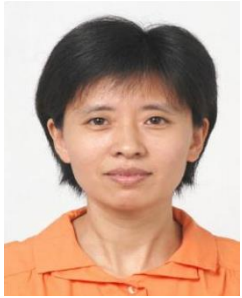

**Shuli He** gained her Ph.D. in material science and engineering from the Northeastern University in China in 1997. She joined the Department of Physics at Capital Normal University in 2000 after postdoc at Zhengzhou University, and was promoted to full professor in 2010. Her research interests include nanoscale magnetism and magnetic materials.

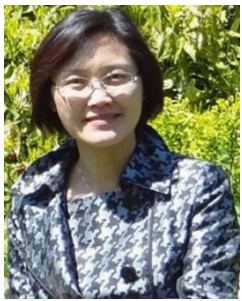

**Xiaohong Xu** received her Ph.D. in Material Science and Engineering from Xi'an Jiaotong University, China in 2001. From 2001 to 2006, she was in Huazhong University of Science and Technology, China, the University of Sheffield, UK, and Tohoku University, Japan as a postdoc or research fellow, In 2002, she become a professor in Material Science at Shanxi Normal University. Her research includes oxide semiconductor spintronics, magnetic recording media, interface physics of heterostructures, two-dimensional materials and localized surface plasmon resonance. She is a Distinguished Young Scholar by the National Science Foundation of China.

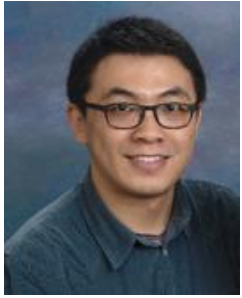

**Qiaoqiang Gan** received his Bachelor's degree from Fudan University, Shanghai, China in 2003, his Master's degree in 2006 in nanophotonics at the Nano Optoelectronics Lab in the Institute of Semiconductors at the Chinese Academy of Sciences, and his Ph.D. degree from Lehigh University in 2010. His current research interests include energy harvesting, conversion, nanophotonics, plasmonics, and bio-photonics. His research publications include over 90 technical papers and 4 patents. He serves as the associate editor for Scientific Reports (NPG), IEEE Photonics Journal and J. of Photonics for Energy (SPIE).

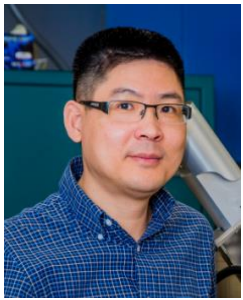

**Hao Zeng** received his Ph.D. in physics from the University of Nebraska-Lincoln in 2001. He joined the Department of Physics at the University at Buffalo in 2004 after a 3-year postdoc at IBM T.J. Watson Research Center, and was promoted to full professor in 2014. His research interests are in nanoscale magnetism and magnetic materials, spintronics, and photovoltaic materials and devices.